\documentclass[twocolumn,prb,showpacs,preprintnumbers,amsmath,amssymb,floatfix]{revtex4}
\usepackage{graphicx}
\usepackage{amsmath}
\usepackage{amssymb}
\usepackage{color}

\begin{document}

\title{Large polaron formation induced by Rashba spin-orbit coupling}
\author{C. Grimaldi}
\affiliation{LPM, Ecole Polytechnique F\'ed\'erale de
Lausanne, Station 17, CH-1015 Lausanne, Switzerland}

\begin{abstract}
Here the electron-phonon Holstein model with Rashba spin-orbit interaction is studied
for a two dimensional square lattice in the adiabatic limit.
It is demonstrated that a delocalized electron at zero spin-orbit coupling
localizes into a large polaron state as soon as the Rashba term is nonzero.
This spin-orbit induced polaron state has localization length inversely proportional to the
Rashba coupling $\gamma$, and it dominates a wide region of the $\gamma$-$\lambda$
phase diagram, where $\lambda$ is the electron-phonon interaction.
\end{abstract}

\maketitle

\section{Introduction}
\label{intro}

Spin manipulation and control is at the core of
spintronics, a technology that uses the spin of the electrons, rather than
their charge, to transfer and/or process information.\cite{awsc,fabian}
The Rashba spin-orbit (SO) coupling arising in materials lacking
structural inversion symmetry\cite{rashba} plays a leading role in this field
because its strength can be tuned by an applied electric field and
by specific material engineering methods. The SO induced lifted spin
degeneracy may then be used in spin filtering devices and spin transistors.

Whether the main effect of SO coupling is limited to the spin splitting
or it is accompanied by substantial modifications in other electronic properties,
which could be detrimental for the spin propagation,
is of course crucial for the functioning of spin-based devices.
In this respect, an important issue calls into play the role of the SO
interaction on the coupling of electrons to the lattice vibrations (phonons).
In particular, a sensible problem is whether the polaron, that is the quasiparticle
composed by the electron and its phonon cloud, is strengthened or weakened by
the Rashba SO interaction.

In previous works, an enhancement of the polaronic character has been obtained
for a two-dimensional (2D) electron gas with linear Rashba coupling for both
short-range (Holstein model Ref.[\onlinecite{holstein}]) and long-range
(Fr\"ohlich model Ref.[\onlinecite{froelich}]) electron-phonon (el-ph)
interactions.\cite{cgf2007a,cgf2007b,grima2008a} On the contrary, a recent calculation on the 2D
tight-binding Holstein-Rashba model on the square lattice has shown that a large
el-ph interaction gets effectively suppressed by the Rashba SO coupling.\cite{berciu}
At present therefore the role of the Rashba SO coupling on the polaron properties
is not clear, and different models and approximations appear to give quite contradicting
results.

In this article the tight-binding Holstein-Rashba model for one electron coupled to adiabatic
phonons is considered and the corresponding non-linear Schr\"odinger equation for the
polaron wave function is solved numerically.
It is shown that, for el-ph couplings such that the electron is delocalized in the zero SO limit,
the Rashba term creates a large polaron state, with polaron localization length inversely
proportional to the SO strength. Furthermore, the small polaron regime appearing at large el-ph
couplings and zero SO gets weakened (or even suppressed) for sufficiently strong SO couplings.
Hence, the Holstein-Rashba polaron is strengthened or weakened by the SO interaction depending on
whether the el-ph coupling is respectively weak or strong, thereby reconciling the different trends
reported in Refs.[\onlinecite{cgf2007a,berciu}] into one single picture.

\begin{figure*}[t]
\protect
\includegraphics[scale=1.78,clip=true]{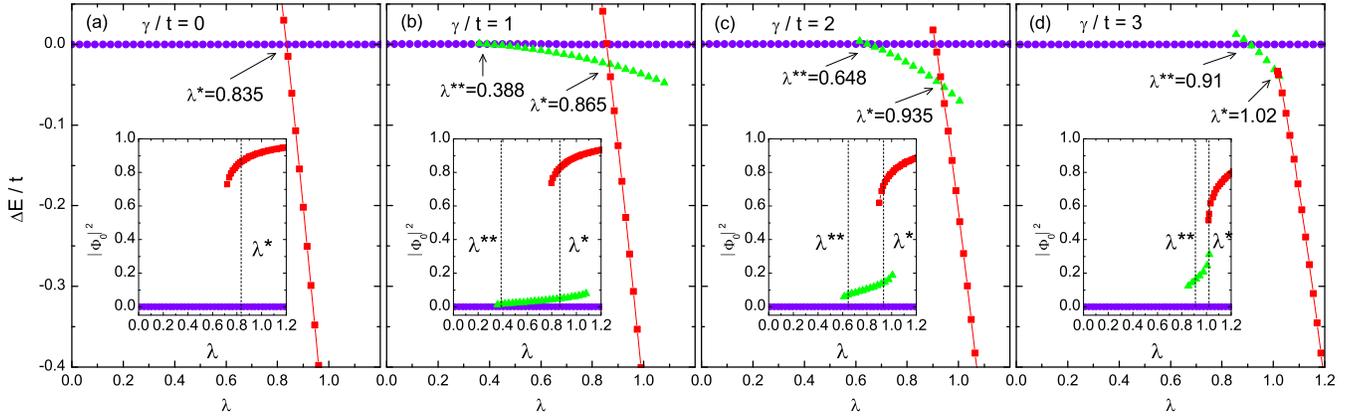}
\caption{(Color online) Total energy difference $\Delta E=E-E_0$ for the adiabatic Holstein-Rashba model
as a function of the
el-ph coupling $\lambda$ and for different values of the SO Rashba interaction $\gamma$. $E_0$ is the
ground state energy for $\lambda=0$. Different symbols refer to different solutions of the nonlinear Schr\"odinger
equation \eqref{nonlin}, and the ground state is given by the solution with lower $\Delta E$ values.
Insets: corresponding electron density probability at $\mathbf{R}=\mathbf{0}$.} \label{fig1}
\end{figure*}

\section{model}
\label{model}

By presenting the spinor operator $\Psi^\dagger_\mathbf{R}=(c^\dagger_{\mathbf{R}\uparrow},c^\dagger_{\mathbf{R}\downarrow})$,
where $c^\dagger_{\mathbf{R}\alpha}$ creates an electron with spin $\alpha=\uparrow,\downarrow$
on site $\mathbf{R}$, the tight-binding Holstein-Rashba Hamiltonian on the square lattice can
be written as  $H=H_0+H_{ph}+H_{el-ph}$, where \cite{sheng}
\begin{align}
\label{h0}
H_0=&-t\sum_\mathbf{R}\left(\Psi^\dagger_\mathbf{R}\Psi_{\mathbf{R}+\hat{\mathbf{x}}}
+\Psi^\dagger_\mathbf{R}\Psi_{\mathbf{R}+\hat{\mathbf{y}}}\right) \nonumber \\
&-i\frac{\gamma}{2}\sum_\mathbf{R}\left(\Psi^\dagger_\mathbf{R}\sigma_y\Psi_{\mathbf{R}+\hat{\mathbf{x}}}
-\Psi^\dagger_\mathbf{R}\sigma_x\Psi_{\mathbf{R}+\hat{\mathbf{y}}}\right)+{\rm H.c.},
\end{align}
is the lattice Hamiltonian for a free electron with transfer integral $t$ and SO coupling $\gamma$.
$\sigma_x$ and $\sigma_y$ are Pauli matrices. The lattice constant is taken to be unity,
and $\hat{\mathbf{x}}$ and $\hat{\mathbf{y}}$ are unit vectors along the $x$ and $y$ directions, respectively.
The Hamiltonian \eqref{h0} is easily diagonalized in momentum space, and the resulting electron
dispersion is composed of two branches: $E_\mathbf{k}^\pm = -2t[\cos(k_x)+\cos(k_y)]
\pm \gamma \sqrt{\sin(k_x)^2+\sin(k_y)^2}$. The lowest branch, $E_\mathbf{k}^-$, has a four-fold
degenerate minimum $E_0=-4t\sqrt{1+\gamma^2/(8t^2)}$ for momenta $\mathbf{k}=(\pm k_0,\pm k_0)$ with
$k_0=\arctan[\gamma/(\sqrt{8}t)]$.\cite{berciu}
The Hamiltonian for Einstein phonons with mass $M$ and frequency $\omega_0$ is given by:
\begin{equation}
\label{hph}
H_{ph}=\sum_\mathbf{R}\left(\frac{P_\mathbf{R}^2}{2M}+\frac{1}{2}M\omega_0^2 X_\mathbf{R}^2\right),
\end{equation}
where $P_\mathbf{R}$ and $X_\mathbf{R}$ are impulse and displacement phonon operators. Finally, the
el-ph Hamiltonian contribution is
\begin{equation}
\label{helph}
H_{el-ph}=\sqrt{2M\omega_0}\,g\sum_\mathbf{R}\Psi^\dagger_\mathbf{R}\Psi_\mathbf{R}X_\mathbf{R},
\end{equation}
where $g$ is the el-ph interaction matrix element.

The (quasi-) 2D materials and heterostructures which display non-zero
Rashba couplings (semiconductor quantum wells, surface states of metals and semimetals)
are wide electron bandwidth systems with $t$ of the order of $1$ eV, while the typical phonon energy
scale is of the order of few to tens meV.\cite{note1} These systems are expected therefore to be
well within the adiabatic regime $\omega_0/t\ll 1$. In the following, however, only the strict
adiabatic limit $\omega_0/t=0$ is considered, which simplifies considerably the problem and, as
shown below, permits to identify the critical parameters governing the electron localization transitions.

The adiabatic limit $\omega_0/t=0$ is obtained formally from Eqs. \eqref{hph} and \eqref{helph} by
setting $M\rightarrow\infty$ and keeping $K=M\omega_0^2$ finite.
Since for $M\rightarrow\infty$ the phonon kinetic energy is zero, the ground state in the adiabatic limit
is obtained by finding the displacement configuration $X_\mathbf{R}^0$ which
minimizes the total energy $E=\langle H \rangle$, where the brackets mean the expectation value
with respect to the electron wave function and the lattice displacement. Hence, since by Hellmann-Feynman theorem $X_\mathbf{R}^0=\sqrt{2M\omega_0}\,g
\langle\psi\vert\Psi^\dagger_\mathbf{R}\Psi_\mathbf{R}\vert\psi\rangle/K$, the ground state energy becomes
\begin{equation}
\label{Hmin}
E_{GS}=\langle\psi\vert H_0\vert\psi\rangle-
E_P\sum_\mathbf{R}\langle\psi\vert\Psi^\dagger_\mathbf{R}\Psi_\mathbf{R}\vert\psi\rangle^2,
\end{equation}
where $E_P=g^2/\omega_0$ is independent of $M$, and $\vert\psi\rangle=\sum_{\mathbf{R},\alpha}\phi_{\mathbf{R}\alpha}c^\dagger_{\mathbf{R}\alpha}\vert 0\rangle$.
The ground state electron wave function $\phi_{\mathbf{R}\alpha}$ can be found from Eq.\eqref{Hmin} by applying the variational
principle, leading to the following non-linear Schr\"odinger equation:
\begin{align}
\label{nonlin}
\varepsilon\Phi_\mathbf{R}=&-t\sum_{n=\pm}(\Phi_{\mathbf{R}+n\hat{\mathbf{x}}}
+\Phi_{\mathbf{R}+n\hat{\mathbf{y}}})-2E_P\mid\Phi_\mathbf{R}\mid^2\Phi_\mathbf{R}\nonumber \\
&-i\frac{\gamma}{2}\sum_{n=\pm}n(\sigma_y \Phi_{\mathbf{R}+n\hat{\mathbf{x}}}
-\sigma_x \Phi_{\mathbf{R}+n\hat{\mathbf{y}}}),
\end{align}
where $\Phi_\mathbf{R}=(\phi^*_{\mathbf{R}\uparrow},\phi^*_{\mathbf{R}\downarrow})^+$ and
$\varepsilon=E_{GS}+E_P\sum_\mathbf{R}\vert\Phi_\mathbf{R}\vert^4$.
Finally, the ground state energy $E_{GS}$ is obtained by solving
Eq.\eqref{nonlin} iteratively, with $\sum_{\mathbf{R},\alpha}\vert\phi_{\mathbf{R}\alpha}\vert^2 = 1$,
and by inserting the resulting wave function into Eq.\eqref{Hmin}.

\section{Results}
\label{results}

Solutions of \eqref{nonlin} for lattices of $N=101\times 101$ sites are plotted in Fig.~\ref{fig1} as a
function of the el-ph coupling constant $\lambda=E_P/(4t)=g^2/(4t\omega_0)$ and for four different
values of $\gamma$. For $\gamma=0$, Fig.~\ref{fig1}(a), we recover the well-known behavior of the adiabatic
Holstein model in two-dimensions:\cite{2D}
a delocalized solution with $E_{\rm GS}=E_0=-4t$ (filled circles)
extending to the whole range of $\lambda$ values considered, and a localized one (filled squares) having
energy lower than $E_0$ for $\lambda\geq \lambda^*=0.835$.
The delocalized/localized nature of the solutions is illustrated in the
inset of Fig.~\ref{fig1}(a) where the electron density probability $\vert\Phi_\mathbf{R}\vert^2=
\sum_\alpha\vert\phi_{\mathbf{R}\alpha}\vert^2$ is plotted for $\mathbf{R}=\mathbf{0}$.
The solution having lower energy for $\lambda\geq \lambda^*$ is a small polaron state, with
more than $90$\% of its wave function localized at the origin.

\begin{figure}[t]
\protect
\includegraphics[scale=0.85,clip=true]{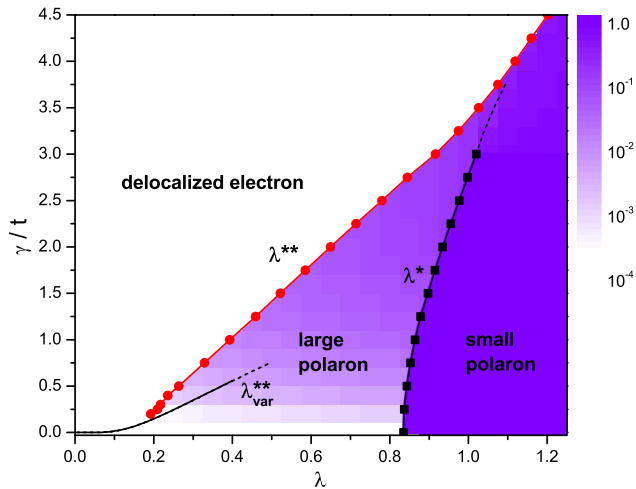}
\caption{(Color online) Phase diagram of the 2D adiabatic Holstein-Rashba model.
The $\lambda^{**}$ and $\lambda^{*}$ transition lines are the phase boundaries separating
the different states of the polaron. The dashed curve has been obtained from the
maximum of $d^2 E/\lambda^2$ and identifies a smooth crossover from large to small
polaron for large $\gamma/t$ values. The solid line is the variational result of Eq.\eqref{var7}.
The graded gray (violet) scale refers to the polaron density probability at $\mathbf{R}=\mathbf{0}$.}\label{fig2}
\end{figure}

Let us now consider the $\gamma>0$ case.
As shown in Figs.~\ref{fig1}(b)-(c), a nonzero Rashba term gives rise to a new feature absent for $\gamma=0$.
Namely, besides the two solutions already discussed for the $\gamma=0$ case,
a third solution appears (filled triangles), which has lower energy than the
delocalized and small polaron states in a region of intermediate values of $\lambda$.
It is thus possible to identify
a second critical coupling, $\lambda^{**}$, such that for $\lambda^{**}\leq \lambda\leq\lambda^*$ the ground state
is given by this third solution. Furthermore, the transition to the small polaron state (identified by $\lambda^*$)
gets shifted to larger el-ph couplings as $\gamma/t$ increases, thereby confirming the results
of Ref.[\onlinecite{berciu}] obtained by a different method and for $\omega_0/t\neq 0$.
A map of the behavior of $\lambda^*$ and $\lambda^{**}$ as $\gamma$ is varied
is reported in the $\gamma/t$-$\lambda$ phase diagram of Fig.~\ref{fig2}, where
the filled circles are the calculated values of $\lambda^{**}$, while the filled squares mark the onset
of the small polaron regime ($\lambda^*$).\cite{note2} The resulting diagram is therefore composed of three
separate regions: a delocalized electron with $E_{GS}=E_0$ for $\gamma/t>\lambda^{**}$ (white region),
a small polaron state for large el-ph couplings ($\lambda>\lambda^*$), and a new ground state
in the region comprised between the $\lambda^{**}$ and $\lambda^*$ lines.

\begin{figure}[t]
\protect
\includegraphics[scale=0.8,clip=true]{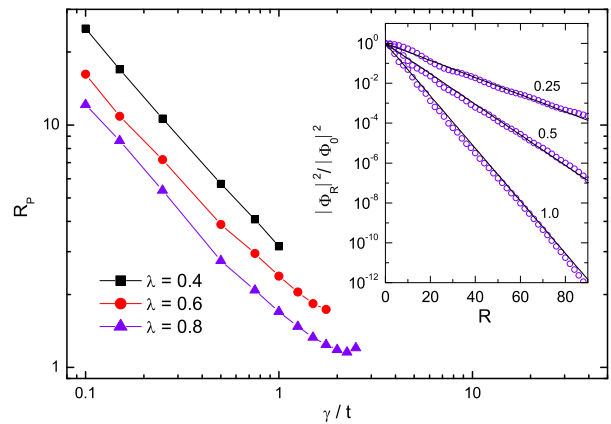}
\caption{(Color online) Polaron radius $R_P$ of the large polaron state as a function of $\gamma/t$
and for different el-ph couplings $\lambda$.
Inset: Density probability (symbols) of the large polaron for $\lambda=0.4$ and $\gamma/t=0.25$, $0.5$, and $1.0$
as a function of distance $R=\vert\mathbf{R}\vert$ along the $(1,0)$ direction.
The solid lines are fits to $\exp(-\vert\mathbf{R}\vert/R_P)$.} \label{fig3}
\end{figure}

As it can be inferred from the insets of Fig.~\ref{fig1}
and from the gray (violet) scale of Fig.~\ref{fig2}, in this region the density probability
at $\mathbf{R}=\mathbf{0}$, $\vert\Phi_\mathbf{0}\vert^2$, is lower than the small polaron solution,
but substantially larger than zero as long as $\gamma\neq 0$, and increasing with $\gamma/t$.
The region between the $\lambda^{**}$ and $\lambda^*$ lines identifies therefore
a large polaron state created by the SO interaction, with a localized wave function which
 may extent over several lattice sites.
The large polaron nature of this solution is substantiated in
Fig.~\ref{fig3}, where the polaron localization radius $R_P$, extracted from a fit of
$\vert\Phi_\mathbf{R}\vert^2$ to $\exp(-\vert\mathbf{R}\vert/R_P)$ (see inset),
is plotted as a function of $\gamma/t$ for $\lambda=0.4$, $0.6$, and $0.8$.
Although a numerical evaluation of $R_P$ for $\gamma/t\rightarrow 0$ is hampered by the finite size
of the lattice, $R_P$ turns out to be approximately
proportional to $t/\gamma$, suggesting therefore that the large polaron evolves continuously
towards a delocalized electron as $\gamma/t\rightarrow 0$.

Further insight on the large polaron state, and in particular on its behavior as $\gamma/t\rightarrow 0$,
can be gained by a simple variational calculation in the continuum. In fact, as long as $R_P$ is
much larger than the lattice constant ($R_P\gg 1$) then an upper bound for $E_{GS}$ can be obtained from
a minimization of the energy functional
\begin{align}
\label{var1}
\mathcal{E}[\Phi(\mathbf{r})]=&\int\! d\mathbf{r}\,\Phi^\dagger(\mathbf{r})\!
\left[t\hat{p}^2+\gamma(\sigma_y\hat{p}_x-
\sigma_x\hat{p}_y)\right]\!\Phi(\mathbf{r})\nonumber\\
&-E_P\int d\mathbf{r}\vert\Phi(\mathbf{r})\vert^4,
\end{align}
where $\hat{p}_q=-i\partial/\partial q$ is the electron momentum operator ($q=x,y$ and
$\hbar=1$) and $\hat{p}^2=\hat{p}_x^2+\hat{p}_y^2$. In the above expression,
$\Phi(\mathbf{r})$ is a suitable ansatz for the ground state spinor, which is assumed to vary
slowly over distances comparable to the lattice spacing. In writing Eq.\eqref{var1},
only the lowest order terms in the lattice constant have been retained,
which amounts to consider a parabolic band with a Rashba coupling linear in the momentum operators.
One can then use for $\Phi(\mathbf{r})$ an ansatz which has been already introduced in studying
the effects of a linear Rashba term on the 2D
Fr\"ohlich polaron and the 2D hydrogen atom:\cite{grima2008a,grima2008b}
\begin{equation}
\label{var2}
\Phi({\bf r})=A\exp(-ar)\left[
\begin{array}{l}
J_0(b r)\\
J_1(b r)\,e^{i\varphi}\!
\end{array}\right].
\end{equation}
Here, $r=\vert\mathbf{r}\vert$ and $\varphi$ is the azimuthal angle, $A$ is a normalization
constant, $J_0$ and $J_1$ are Bessel functions, and $a$ and $b$
are variational parameters.
By using \eqref{var2} and the properties of the Bessel functions, equation \eqref{var1}
reduces to
\begin{align}
\label{var3}
\mathcal{E}&=t(a^2+b^2)-\gamma b-\frac{E_P}{2\pi}\frac{\int_0^\infty\!dr\, r e^{-4ar}F(br)^2}
{\left[\int_0^\infty\!dr\, r e^{-2ar}F(br)\right]^2} \nonumber \\
&\simeq t(a^2+b^2)-\gamma b -\frac{2E_P a^2}{\pi}\ln\left(\frac{b}{\sqrt{e}a}\right),
\end{align}
where $F(br)=J_0(br)^2+J_1(br)^2$. The second equality stems from assuming $a\ll b$,
which is the relevant limit of the large polaron regime. Minimization of $\mathcal{E}$ with
respect to $a$ and $b$ leads to two possible solutions: $b=\gamma/(2t)$ and $a=0$, which corresponds
to a delocalized electron with $\mathcal{E}_{\rm min}=E_0=-4t-\gamma^2/(4t)$, and
\begin{equation}
\label{var4}
a=b\exp\left(-1-\frac{\pi}{8\lambda}\right),
\,\,b=\frac{\gamma/(2t)}{1-4\lambda\exp\left[-2-\pi/(4\lambda)\right]/\pi},
\end{equation}
which represents the large polaron solution with
\begin{equation}
\label{var5}
\mathcal{E}_{\rm min}-E_0=
-\frac{\lambda}{\pi}\frac{\gamma^2}{t}\exp\left(-2-\frac{\pi}{4\lambda}\right),
\end{equation}
for $\lambda$ small. Since Eq.\eqref{var5} is an upper bound for $\Delta E_{\rm GS}=E_{\rm GS}-E_0$,
then that the large polaron state has energy always lower than the delocalized electron.
Furthermore, by realizing that the variational parameter $a$ represents the polaron radius through $a=1/(2R_P)$,
it turns out from Eq.\eqref{var4} that $R_P$ scales as $t/\gamma$, in agreement therefore with the
results of Fig.~\ref{fig3}.

The finding that a large polaron is formed for $\gamma/t\neq 0$ is in accord with the observation
of Ref.[\onlinecite{cgf2007a}] that perturbation theory breaks down in the adiabatic limit for
any finite $\lambda$. This breakdown basically stems from the one-dimensional-like divergence of
the density of states (DOS) of a parabolic band with linear Rashba coupling.\cite{cgf2007a,cgf2007b}

Although the variational result presented above correctly predicts the appearance of the large polaron state
as soon as $\gamma/t\neq 0$, it fails nevertheless in describing the $\lambda^{**}$
transition line of Fig.~\ref{fig2} separating the large polaron state from the delocalized solution.
This is because the lowest order expansion in the lattice constant of Eq.~\eqref{var1} neglects
higher order powers of the momentum operator arising from the lattice Rashba term, which
shift the van Hove divergence of the DOS from $E_0$ to higher energies,\cite{berciu}
thereby making the perturbation theory non-singular.
To investigate this point within the variational method, it suffices to
expand the discrete Hamiltonian up to the third order in the lattice constant. This corresponds
to add to the energy functional \eqref{var1} the following contribution
\begin{equation}
\label{var6}
\mathcal{E}'[\Phi(\mathbf{R})]=\frac{\gamma}{6}\int\! d\mathbf{r}\,\Phi^\dagger(\mathbf{r})\!
\left(\sigma_x\hat{p}_y^3-\sigma_y\hat{p}_x^3\right)\!\Phi(\mathbf{r}),
\end{equation}
which, by using again the ansatz \eqref{var2} and for $a\ll b$, leads to the third order correction term
$\mathcal{E}'=(\gamma/8)(b^3+3a^2b-\pi a^3)$ to Eq.\eqref{var3}. It is then easy
to shown that $[\mathcal{E}+\mathcal{E}']_{\rm min}-E_0$ is negative
[with $E_0=-4t-\gamma^2/(4t)+\gamma^4/(128 t^3)$] as long as $\gamma/t< \lambda^{**}_{\rm var}$, where
for $\lambda$ small
\begin{equation}
\label{var7}
\lambda^{**}_{\rm var}=8\sqrt{\frac{2\lambda}{\pi}}\exp\!\left(-1-\frac{\pi}{8\lambda}\right).
\end{equation}
Although Eq.~\eqref{var7} provides only a lower bound for
$\lambda^{**}$ (solid line in Fig.~\ref{fig2}), it shows nevertheless that, as $\gamma/t$ is enhanced for
fixed $\lambda$, the transition from the large polaron to the delocalized electron state
originates from higher order of the SO interaction than the linear Rashba coupling.

\section{Discussion and conclusions}
\label{concl}

Let us discuss now the significance of the results reported above for materials of interest and
possible consequences for spintronics applications. First of all, it is important to identify the
region in the phase diagram of Fig.~\ref{fig2} where realistic values of $\gamma/t$ and $\lambda$
are expected to fall. This is easily done by realizing that the largest Rashba SO coupling to date
is that found in the surface stats of Bi/Ag(111) surface alloys\cite{ast} for which
$\gamma/t\approx 1.4$ can be estimated. Other 2D systems and heterostructures have lower or much lower
$\gamma/t$ values. Concerning the coupling to the phonons, a survey\cite{kroger} on the el-ph interaction
at metal surfaces evidences that $\lambda$ is usually lower than $0.6$-$0.7$ (see also Ref.[\onlinecite{hofmann}]),
at least for the surface states with large SO splittings (\emph{i.e.} Ag, Cu, Bi). It is therefore
a rather conserving assumption to confine to $\gamma/t\lesssim 1$ and $\lambda \lesssim 1$ the region
of interest for the microscopic parameters which, as shown in Fig.~\ref{fig2}, is substantially
dominated by the SO induced large polaron state.
Hence, upon tuning of the Rashba SO coupling, a delocalized electron at $\gamma/t=0$ can in principle
be changed into a self-trapped large polaron state for $\gamma/t>0$, with obvious consequences on the
spin propagation in the system. In passing, it is worth noticing
that the small polaron regime instead is affected rather weakly by the SO interaction for $\gamma/t \lesssim 1$,
while its weakening gets pronounced only for unrealistically large values of
$\gamma/t$ (see also Fig.~\ref{fig1}).

Before concluding, it is important to discuss a last important point. Although the adiabatic limit
employed here allows for a clear identification of the $\lambda^*$ and $\lambda^{**}$ transition lines,
the energy gain associated to the large polaron formation becomes very small in the weak coupling and
small SO limits [see Eq.~\eqref{var5}]. In this regime, the inclusion of quantum fluctuations which
arise as soon as $\omega_0/t\neq 0$ may wash out completely any signature (like e.g. an anomalous
enhancement of the electron effective mass $m^*$) of the large polaron state,
even for $\omega_0/t$ small, while they should remain visible for larger $\lambda$ and $\gamma/t$
values.  For a more complete description of the SO effects on the Holstein-Rashba polaron, it is
therefore necessary to extend the study to the non-adiabatic regime  $\omega_0/t\neq 0$, by keeping
however in mind that, as discussed above, relevant materials have $\omega_0/t \ll 1$.

In summary, the complete phase diagram of the 2D adiabatic Holstein el-ph Hamiltonian in the
presence of Rashba SO coupling has been calculated.
It has been shown that a self-trapped large polaron state is created by the SO
interaction in a wide region of the phase diagram, and that its localization radius can be modulated by the
SO coupling. This result implies that, for realistic values of the microscopic parameters, the appearance
of a self-trapped large polaron state is a potentially detrimental factor for spin transport.

\acknowledgements
The author thanks E. Cappelluti, S. Ciuchi, and F. Marsiglio for valuable comments.


\begin{thebibliography}{99}
\vskip -0.5cm

\bibitem{awsc}
D. Awschalom and N. Samarth, Physics {\bf2}, 50 (2009).

\bibitem{fabian}
I. \v{Z}uti\'{c}, J. Fabian, and S. Das Sarma,
Rev. Mod. Phys. {\bf 76}, 323 (2004).

\bibitem{rashba}
E. I. Rashba, Sov. Phys. Solid State {\bf 2}, 1109 (1960).

\bibitem{holstein}
T. Holstein, Ann. Phys. {\bf 8}, 325 (1959); {\bf 8}, 343 (1959).

\bibitem{froelich}
H. Fr\"ohlich, Adv. Phys. {\bf 3}, 325 (1954).

\bibitem{cgf2007a}
E. Cappelluti, C. Grimaldi, and F. Marsiglio, Phys. Rev. B {\bf 76}, 085334 (2007).

\bibitem{cgf2007b}
E. Cappelluti, C. Grimaldi, and F. Marsiglio, Phys. Rev. Lett.
{\bf 98}, 167002 (2007)..

\bibitem{grima2008a}
C. Grimaldi, Phys. Rev. B {\bf 77}, 024306 (2008).

\bibitem{berciu}
L. Covaci and M. Berciu, Phys. Rev. Lett. {\bf 102}, 186403 (2009).

\bibitem{sheng}
L. Sheng, D. N. Sheng, and C. S. Ting, Phys. Rev. Lett. {\bf 94}, 016602 (2005).

\bibitem{note1}
It should be notes also that large SO splittings are expected in systems
whose constituting elements have large atomic
number $Z$, and so large mass number. As a rule of thumb therefore, larger values of
$\gamma$ are accompanied by lower phonon frequencies $\omega_0$.

\bibitem{2D}
A. Lagendijk and H. De Raedt, Phys. Lett. A {\bf 108}, 91 (1985);
V. V. Kabanov and O. Yu. Mashtakov, Phys. Rev. B {\bf 47}, 6060 (1993).

\bibitem{note2}
Despite that lattices up to $1001\times 1001$ sites have been considered in compiling Fig.~\ref{fig2},
it has not been possible to identify with sufficient accuracy the delocalized
electron / large polaron transition line $\lambda^{**}$ for $\gamma/t<0.2$, because of the tiny energy
differences involved.

\bibitem{grima2008b}
C. Grimaldi, Phys. Rev. B {\bf 77}, 113308 (2008).

\bibitem{ast}
C. R. Ast, J. Henk, A. Ernst, L. Moreschini, M. C. Falub,
D. Pacil\'e, P. Bruno, K. Kern, and M. Grioni, Phys. Rev. Lett. {\bf 98}, 186807 (2007).

\bibitem{kroger}
J. Kr\"oger, Rep. Prog. Phys. {\bf 69}, 899 (2006).

\bibitem{hofmann}
Ph. Hofmann, Prog. Surf. Sci. {\bf 81}, 191 (2006).

\end{thebibliography}
\end{document}